\documentclass{elsart}
\usepackage{ifpdf}
\usepackage{graphicx,amssymb,lineno,epsfig,subfig}
\usepackage{subfig}

\usepackage{indentfirst}

\ifpdf
\usepackage[%
  pdftitle={Instructions for use of the document class
    elsart},%
  pdfauthor={Simon Pepping},%
  pdfsubject={The preprint document class elsart},%
  pdfkeywords={instructions for use, elsart, document class},%
  pdfstartview=FitH,%
  bookmarks=true,%
  bookmarksopen=true,%
  breaklinks=true,%
  colorlinks=true,%
  linkcolor=blue,anchorcolor=blue,%
  citecolor=blue,filecolor=blue,%
  menucolor=blue,pagecolor=blue,%
  urlcolor=blue]{hyperref}
\else
\usepackage[%
  breaklinks=true,%
  colorlinks=true,%
  linkcolor=blue,anchorcolor=blue,%
  citecolor=blue,filecolor=blue,%
  menucolor=blue,pagecolor=blue,%
  urlcolor=blue]{hyperref}
\fi

\renewcommand{\theequation}{\arabic{section}.\arabic{equation}}

\def\@subjclass{}
\makeatletter
\def\elsartstyle{%
    \def\normalsize{\@setfontsize\normalsize\@xiipt{14.5}}
    \def\small{\@setfontsize\small\@xipt{13.6}}
    \let\footnotesize=\small
    \def\large{\@setfontsize\large\@xivpt{18}}
    \def\Large{\@setfontsize\Large\@xviipt{22}}
    \skip\@mpfootins = 18\p@ \@plus 2\p@
    \normalsize
} \@ifundefined{square}{}{\let\Box\square} \makeatother
\usepackage{indentfirst}
\usepackage{setspace}
\begin{document}
\begin{frontmatter}
\title{Periodic and solitary wave solutions to the Fornberg-Whitham equation}
\author{Jiangbo Zhou\corauthref{cor}},
\corauth[cor]{Corresponding author. } \ead{zhoujiangbo@yahoo.cn}
\author{Lixin Tian}
\address{Nonlinear Scientific Research Center, Faculty of Science, Jiangsu
University, Zhenjiang, Jiangsu 212013, China}
\begin{abstract} In this paper, new travelling wave solutions to the
Fornberg-Whitham equation
 \[u_t - u_{xxt} + u_x + uu_x= uu_{xxx}+
3u_x u_{xx} \]
 are investigated. They are characterized by
two parameters. The expresssions of the periodic and solitary wave
solutions are obtained.
\end{abstract}


\end{frontmatter}
\section{Introduction}
 \setcounter {equation}{0}
Recently, Ivanov \cite{1} investigated the integrability of a class
of nonlinear dispersive wave equations
\begin{equation}
\label {eq1.1}u_t - u_{xxt} + \partial_x (\kappa u+\alpha u^2+\beta
u^3)=\nu u_x u_{xx}+\gamma uu_{xxx},
\end{equation}
where and $\alpha, \beta, \gamma, \kappa, \nu$  are real constants.

 The important cases of Eq.(\ref{eq1.1}) are:

 The hyperelastic-rod wave equation
\begin{equation}
\label {eq1.2}u_t - u_{xxt} + 3uu_x =\gamma(2u_x u_{xx}+ uu_{xxx}),
\end{equation}
has been recently studied as a model, describing nonlinear
dispersive waves in cylindrical compressible hyperelastic rods
\cite{2}-\cite{7}. The physical parameters of various compressible
materials put $\gamma$ in the range from -29.4760 to 3.4174
\cite{2,4}.

The Camassa-Holm equation
\begin{equation}
\label{eq1.3} u_t - u_{xxt} + 3uu_x = 2u_x u_{xx} + uu_{xxx},
\end{equation}
describes the unidirectional propagation of shallow water waves over
a flat bottom \cite{8,9}. It is completely integrable \cite{1} and
admits, in addition to smooth waves, a multitude of travelling wave
solutions with singularities: peakons, cuspons, stumpons and
composite waves \cite{9}-\cite{12}. The solitary waves of
Eq.(\ref{eq1.2}) are smooth if $\kappa>0$ and peaked if $\kappa=0$
\cite{9,10}. Its solitary waves are stable solitons \cite{13,14},
retaining their shape and form after interactions \cite{15}. It
models wave breaking \cite{16}-\cite{18}.

The Degasperis-Procesi equation
\begin{equation}
\label{eq1.4} u_t - u_{xxt} + 4uu_x = 3u_x u_{xx} + uu_{xxx},
\end{equation}
models nonlinear shallow water dynamics. It is completely integrable
\cite{1} and has a variety of
 travelling wave solutions including solitary wave solutions, peakon solutions and shock
 waves solutions \cite {20}-\cite {27}.

The Fornberg-Whitham equation
\begin{equation}
\label {eq1.5}u_t - u_{xxt} + u_x + uu_x= uu_{xxx}+ 3u_x u_{xx},
\end{equation}
appeared in the study qualitative behaviors of wave-breaking \cite
{29}. It admits a wave of greatest height, as a peaked limiting form
of the travelling wave solution \cite {30}, $u(x,t) = A\exp
(-\frac{1}{2}\left| {x - \frac{4}{3}t} \right|)$, where $A$ is an
arbitrary constant. It is not completely integrable \cite{1}.

The regularized long-wave or BBM equation

\begin{equation}
\label{eq1.6} u_t - u_{xxt} + u_x +uu_x = 0,
\end{equation}
and the modified BBM equation
\begin{equation}
\label{eq1.7} u_t - u_{xxt} + u_x +3u^2u_x = 0,
\end{equation}
have also been investigated by many authors \cite{31}-\cite{39}.

Many efforts have been devoted to study
Eq.(\ref{eq1.2})-(\ref{eq1.4}),(\ref{eq1.6}) and (\ref{eq1.7}),
however, little attention was paid to study Eq.(\ref{eq1.5}). In
\cite {40}, we constructed two types of bounded travelling wave
solutions $u(\xi) (\xi=x-ct)$ to Eq.(\ref{eq1.5}), which are defined
on semifinal bounded domains and called kink-like and antikink-like
wave solutions. In this paper, we continue to study the travelling
wave solutions to Eq.(\ref{eq1.5}). Following
 Vakhnenko and Parkes's strategy in \cite{19}, we obtain some
 periodic and solitary wave solutions $u(\xi)$ to
Eq.(\ref{eq1.5}) which are defined on $(-\infty, +\infty)$. The
travelling wave solutions obtained in this paper are obviously
different from those obtained in our previous work \cite {40}. To
the best of our knowledge, these solutions are new for
Eq.(\ref{eq1.5}). Our work may help people to know deeply the
described physical process and possible applications of the
Fornberg-Whitham equation.

The remainder of the paper is organized as follows. In Section 2,
for completeness and readability, we repeat Appendix A in \cite{19},
which discuss the solutions to a first-order ordinary differential
equaion. In Section 3, we show that, for travelling wave solutions,
Eq.(\ref{eq1.5}) may be reduced to a first-order ordinary
differential equation involving two arbitrary integration constants
$a$ and $b$. We show that there are four distinct periodic solutions
corresponding to four different ranges of values of $a$ and
restricted ranges of values of $b$. A short conclusion is given in
Section 4.

\section{Solutions to a first-order ordinary differential
equaion}
 \setcounter {equation}{0}

This section is due to Vakhnenko and Parkes (see Appendix A in
\cite{19}). For completeness and readability, we repeat it in the
following.

Consider solutions to the following ordinary
 differential equation
\begin{equation}
\label{eq2.1}
 (\varphi\varphi_\xi)^2=\varepsilon^2 f(\varphi),
\end{equation}
where
\begin{equation}
\label{eq2.2}
f(\varphi)=(\varphi-\varphi_1)(\varphi-\varphi_2)(\varphi_3-\varphi)(\varphi_4-\varphi),
\end{equation}
and $\varphi_1$, $\varphi_2$, $\varphi_3$, $\varphi_4$ are chosen to
be real constants with $\varphi_1\leq \varphi_2\leq \varphi\leq
\varphi_3 \leq \varphi_4$.

Following \cite{42} we introduce $\zeta$ defined by
\begin{equation}
\label{eq2.3} \frac{d\xi}{d\zeta}=\frac{\varphi}{\varepsilon},
\end{equation}
so that Eq.(\ref{eq2.1}) becomes
\begin{equation}
\label{eq2.4} (\varphi_\zeta)^2=f(\varphi).
\end{equation}

Eq.(\ref{eq2.4}) has two possible forms of solution. The first form
is found using result 254.00 in \cite{43}. Its parametric form is
\begin{equation}
\label{eq2.5} \left\{ {\begin{array}{l}
\varphi = \frac{\textstyle \varphi_2-\varphi_1 n \mathrm{sn}^2(w|m) }{\textstyle 1-n \mathrm{sn}^2(w|m)} ,\\
 \xi = \frac{\displaystyle 1}{\displaystyle \varepsilon
 p}(w\varphi_1+(\varphi_2-\varphi_1)\Pi(n;w|m)),
\\
 \end{array}} \right.
\end{equation}
with $w$ as the parameter, where
\begin{equation}
\label{eq2.6}
m=\frac{(\varphi_3-\varphi_2)(\varphi_4-\varphi_1)}{(\varphi_4-\varphi_2)(\varphi_3-\varphi_1)},
p=\frac{1}{2}\sqrt{(\varphi_4-\varphi_2)(\varphi_3-\varphi_1)},
w=p\zeta,
\end{equation}
and
\begin{equation}
\label{eq2.7} n=\frac{\varphi_3-\varphi_2}{\varphi_3-\varphi_1}.
\end{equation}
In (\ref{eq2.5}) $\mathrm{sn}(w|m)$ is a Jacobian elliptic function,
where the notation is as used in Chapter 16 of \cite{44}.
$\Pi(n;w|m)$ is the elliptic integral of the third kind and the
notation is as used in Section 17.2.15 of \cite{44}.

The solution to (\ref{eq2.1}) is given in parametric form by
(\ref{eq2.5}) with $w$ as the parameter. With respect to $w$,
$\varphi$ in (\ref{eq2.5}) is periodic with period $2K(m)$, where
$K(m)$ is the complete elliptic integral of the first kind. It
follows from (\ref{eq2.5}) that the wavelength $\lambda$ of the
solution to (\ref{eq2.1}) is
\begin{equation}
\label{eq2.8} \lambda = \frac{\displaystyle 2}{\displaystyle
 \varepsilon p}|\varphi_1K(m)+(\varphi_2-\varphi_1)\Pi(n|m)|,
 \end{equation}
where $\Pi(n|m)$ is the complete elliptic integral of the third
kind.

When $\varphi_3=\varphi_4$, $m=1$, (\ref{eq2.5}) becomes
\begin{equation}
\label{eq2.9} \left\{ {\begin{array}{l}
\varphi = \frac{\textstyle \varphi_2-\varphi_1 n \tanh^2w }{\textstyle 1-n  \tanh^2w} ,\\
 \xi = \frac{\textstyle 1}{\textstyle \varepsilon
 }(\frac{\textstyle w\varphi_3}{\textstyle p}-2\tanh^{-1}(\sqrt{n}\tanh
 w)).
\\
 \end{array}} \right.
\end{equation}

The second form of solution of (\ref{eq2.5}) is found using result
255.00 in \cite{43}. Its parametric form is

\begin{equation}
\label{eq2.10}  \left\{ {\begin{array}{l}
\varphi = \frac{\textstyle \varphi_3-\varphi_4 n \mathrm{sn}^2(w|m) }{\textstyle 1-n \mathrm{sn}^2(w|m)} ,\\
 \xi = \frac{\displaystyle 1}{\displaystyle \varepsilon
 p}(w\varphi_4-(\varphi_4-\varphi_3)\Pi(n;w|m)),
\\
 \end{array}} \right.
\end{equation}
where $m, p, w$ are as in (\ref{eq2.6}), and
\begin{equation}
\label{eq2.11} n=\frac{\varphi_3-\varphi_2}{\varphi_4-\varphi_2}.
\end{equation}

The solution to (\ref{eq2.1}) is given in parametric form by
(\ref{eq2.10}) with $w$ as the parameter. The wavelength $\lambda$
of the solution to (\ref{eq2.1}) is
\begin{equation}
\label{eq2.12} \lambda = \frac{\displaystyle 2}{\displaystyle
 \varepsilon p}|\varphi_4K(m)-(\varphi_4-\varphi_3)\Pi(n|m)|.
 \end{equation}

When $\varphi_1=\varphi_2$, $m=1$, (\ref{eq2.10}) becomes
\begin{equation}
\label{eq2.13} \left\{ {\begin{array}{l}
\varphi = \frac{\textstyle \varphi_3-\varphi_4 n \tanh^2w }{\textstyle 1-n  \tanh^2w} ,\\
 \xi = \frac{\textstyle 1}{\textstyle \varepsilon
 }(\frac{\textstyle w\varphi_2}{\textstyle p}+2\tanh^{-1}(\sqrt{n}\tanh
 w)).
\\
 \end{array}} \right.
\end{equation}

\section{Periodic and solitary wave solutions to Eq.(\ref{eq1.5})}
 \setcounter {equation}{0}
Eq.(\ref{eq1.5}) can also be written in the form
\begin{equation}
\label{eq3.1} (u_t+uu_x)_{xx}= u_t+uu_x+u_x.
 \end{equation}
Let $u=\varphi(\xi )+c$ with $\xi = x - ct$ be a travelling wave
solution to Eq.(\ref{eq3.1}), then it follows that
\begin{equation}
\label{eq3.2} (\varphi \varphi_\xi)_{\xi\xi}= \varphi
\varphi_\xi+\varphi_\xi.
 \end{equation}
Integrating (\ref{eq3.2}) twice with respect to $\xi$, we have
 \begin{equation}
\label{eq3.3}
(\varphi\varphi_\xi)^2=\frac{1}{4}(\varphi^4+\frac{8}{3}\varphi^3+a
\varphi^2+b),
 \end{equation}
where $a$ and $b$ are two arbitrary integration constants.

Eq.(\ref{eq3.3}) is in the form of Eq.(\ref{eq2.1}) with
$\varepsilon=\frac{\textstyle1}{\textstyle2}$ and
$f(\varphi)=(\varphi^4+\frac{\textstyle8}{\textstyle3}\varphi^3+a
\varphi^2+b)$. For convenience we define $g(\varphi)$ and
$h(\varphi)$ by
 \begin{equation}
\label{eq3.4} f(\varphi)=\varphi^2g(\varphi)+b, \  \mbox{where} \
g(\varphi)=\varphi^2+\frac{8}{3}\varphi+a,
 \end{equation}
 \begin{equation}
\label{eq3.5} f'(\varphi)=2\varphi h(\varphi), \ \mbox{where} \
h(\varphi)=2\varphi^2+4\varphi+a,
 \end{equation}
and define $\varphi_L$, $\varphi_R$, $b_L$, and $b_R$ by
 \begin{equation}
\label{eq3.6} \varphi_L=-\frac{\textstyle 1}{\textstyle
2}(2+\sqrt{4-2a}), \varphi_R=-\frac{\textstyle 1}{\textstyle
2}(2-\sqrt{4-2a}),
 \end{equation}
 \begin{equation}
\label{eq3.7} b_L=-\varphi_L^2g(\varphi_L)=\frac{\textstyle
a^2}{\textstyle 4}-2a+\frac{\textstyle 8}{\textstyle
3}+\frac{\textstyle 2}{\textstyle 3}(2-a)\sqrt{4-2a},
 \end{equation}
 \begin{equation}
 \label{eq3.8} b_R=-\varphi_R^2g(\varphi_R)=\frac{\textstyle
a^2}{\textstyle 4}-2a+\frac{\textstyle 8}{\textstyle
3}-\frac{\textstyle 2}{\textstyle 3}(2-a)\sqrt{4-2a}.
 \end{equation}
 Obviously, $\varphi_L$, $\varphi_R$ are the roots of $h(\varphi)=0$.

 In the following, suppose that $a<2$ and $a\neq 0$ such that
 $f(\varphi)$ has three distinct stationary points: $\varphi_L$,
 $\varphi_R$, $0$ and comprise two minimums separated by a maximum.
 Under this assumption, Eq.(\ref{eq3.3}) has periodic and solitary wave
solutions that have different analytical forms depending on the
values of $a$ and $b$ as follows:

(1) $a<0$

In this case $\varphi_L<0<\varphi_R$ and
$f(\varphi_L)<f(\varphi_R)$. For each value $a<0$ and $0<b<b_R$ (a
corresponding curve of $f(\varphi)$ is shown in Fig.1(a)), there are
periodic inverted loop-like solutions to Eq.(\ref{eq3.3}) given by
(\ref{eq2.5}) so that $0<m<1$, and with wavelength given by
(\ref{eq2.8}); see Fig.2(a) for an example.
\begin{figure}[h]
\centering \subfloat[]{\label{fig:1}
\includegraphics[height=1.1in,width=1.2in]{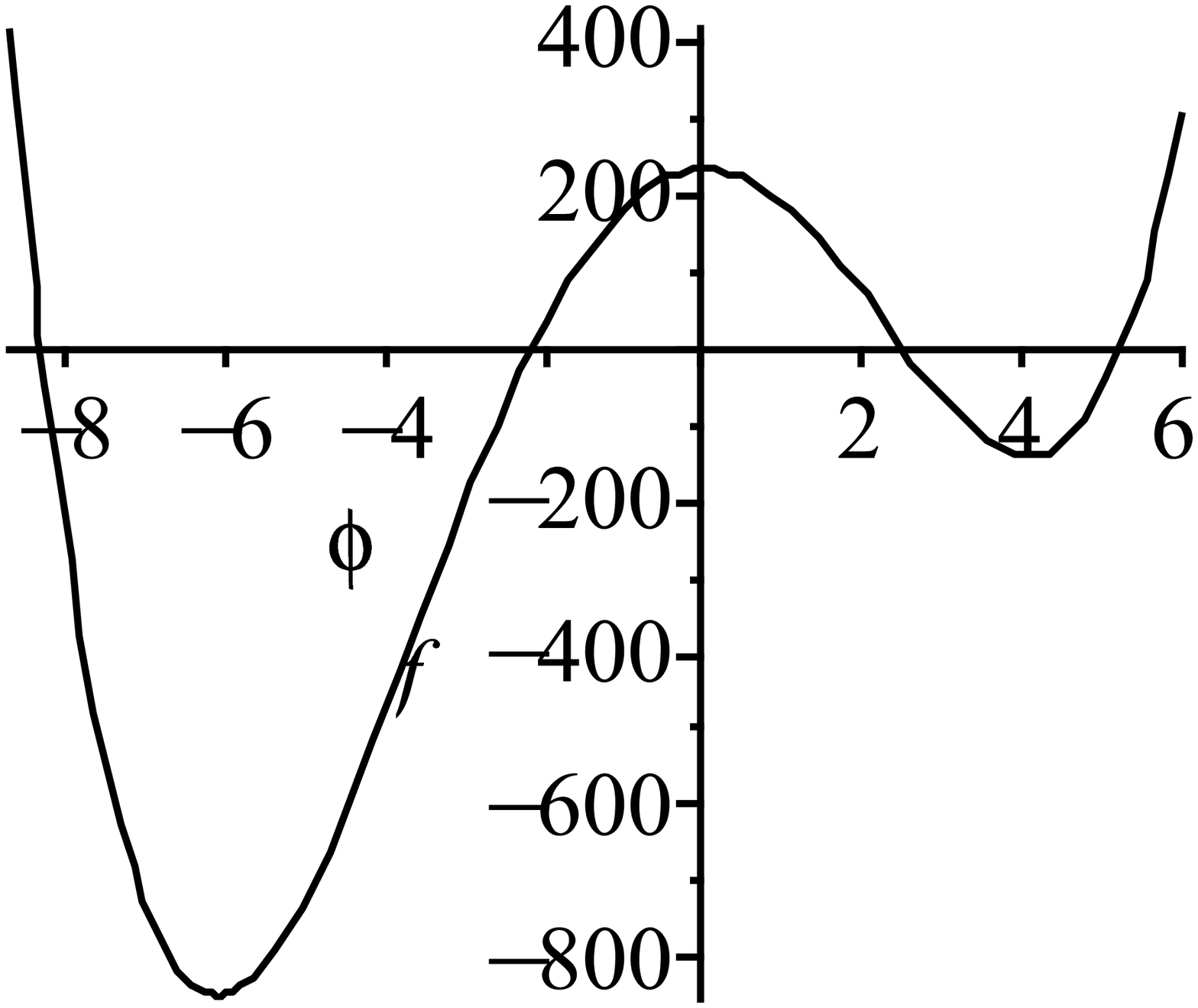}}\hspace{0.01\textwidth}
\subfloat[ ]{ \label{fig:2}
\includegraphics[height=1.1in,width=1.2in]{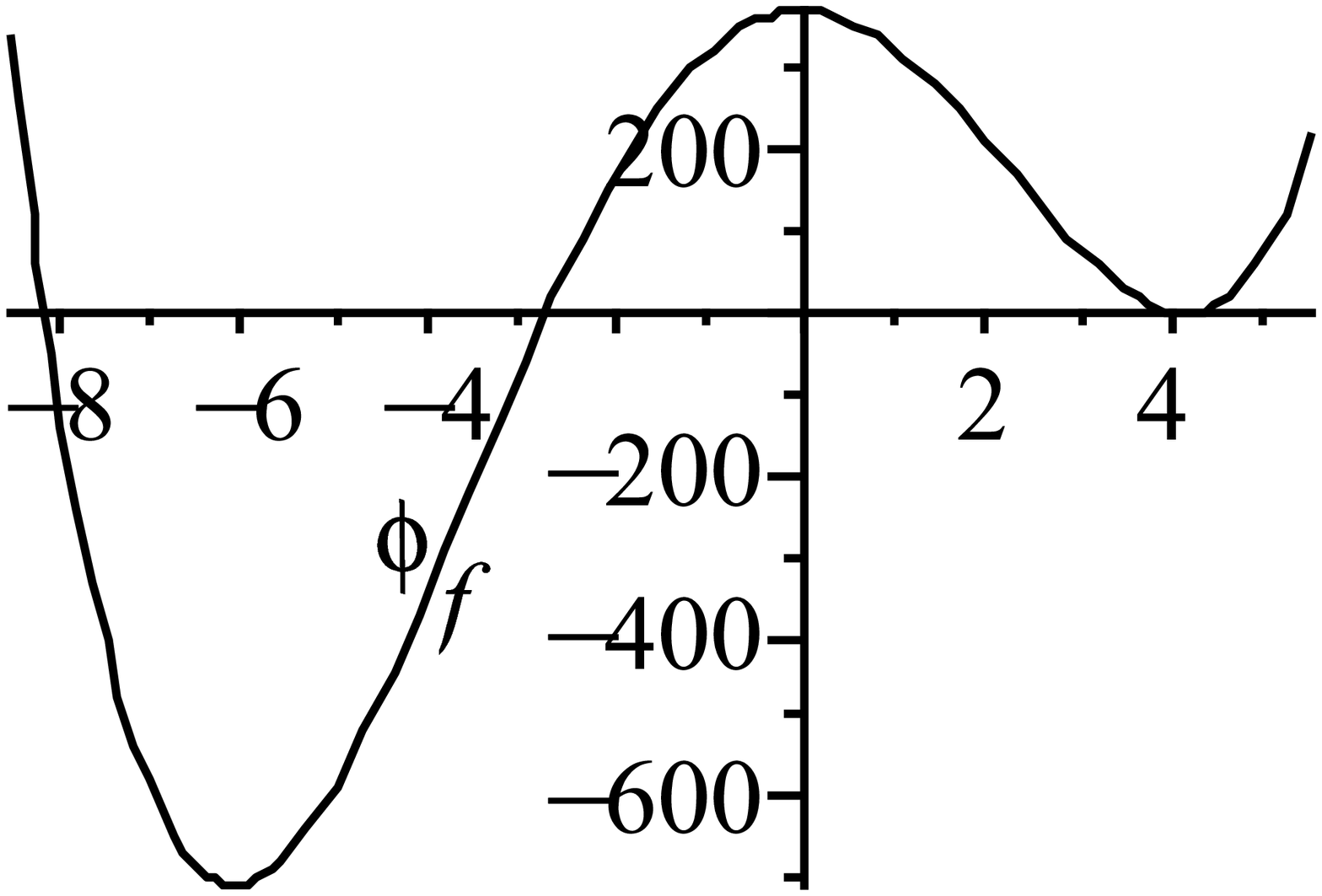}}\hspace{0.01\textwidth}
\subfloat[]{ \label{fig:3}
\includegraphics[height=1.1in,width=1.2in]{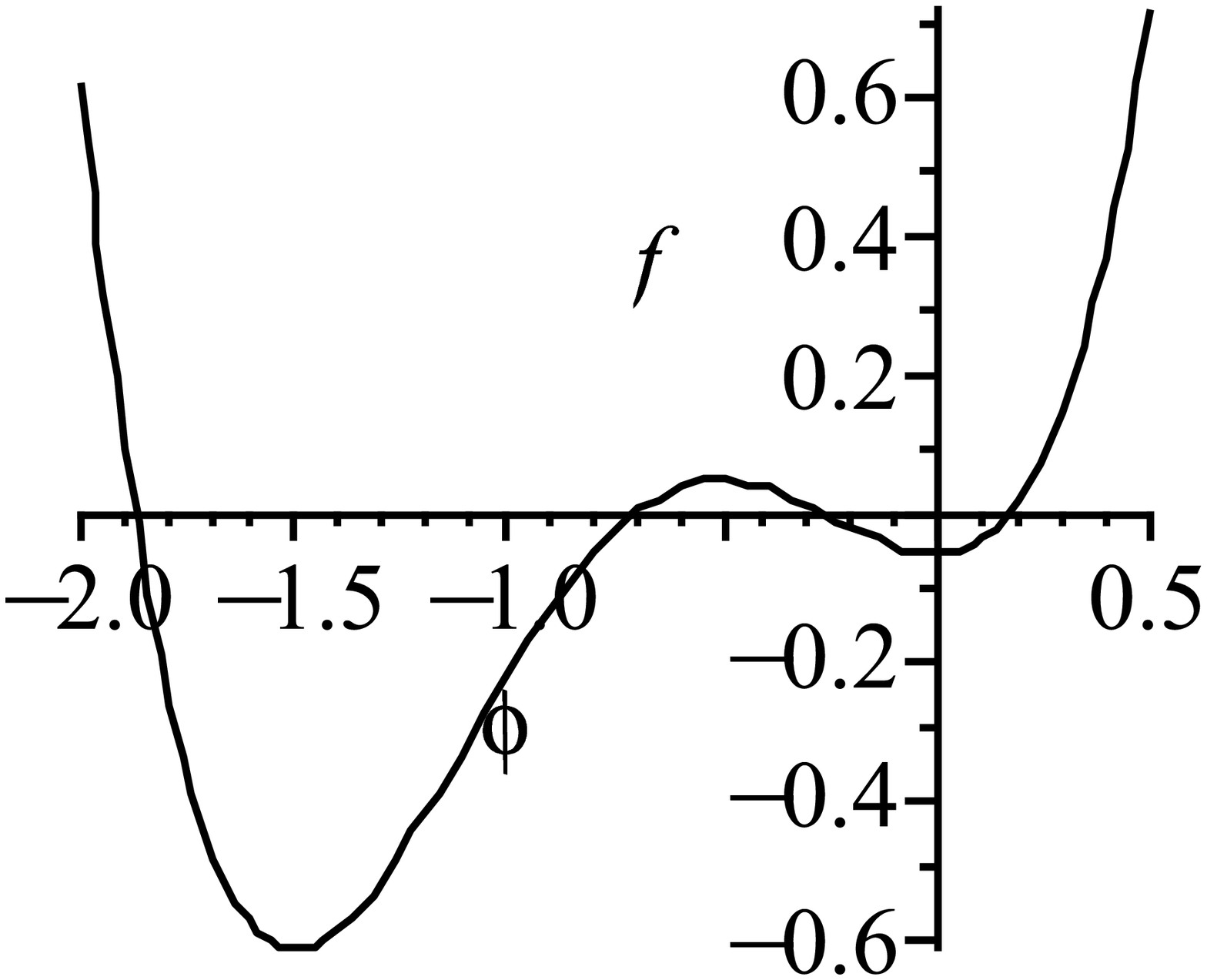}}\hspace{0.01\textwidth}
\subfloat[ ]{ \label{fig:4}
\includegraphics[height=1.1in,width=1.2in]{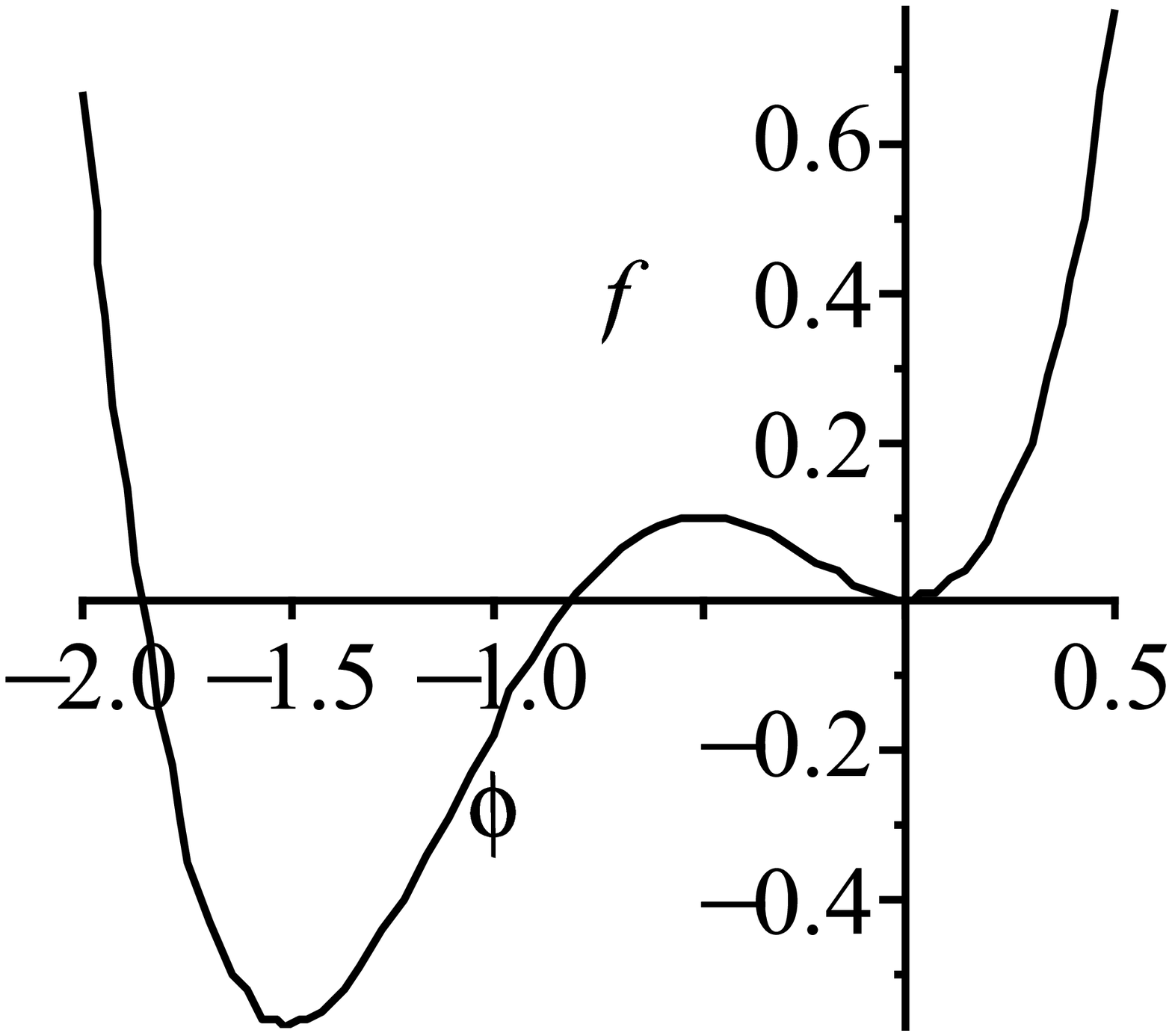}}\\
\subfloat[]{ \label{fig:5}
\includegraphics[height=1.in,width=1.2in]{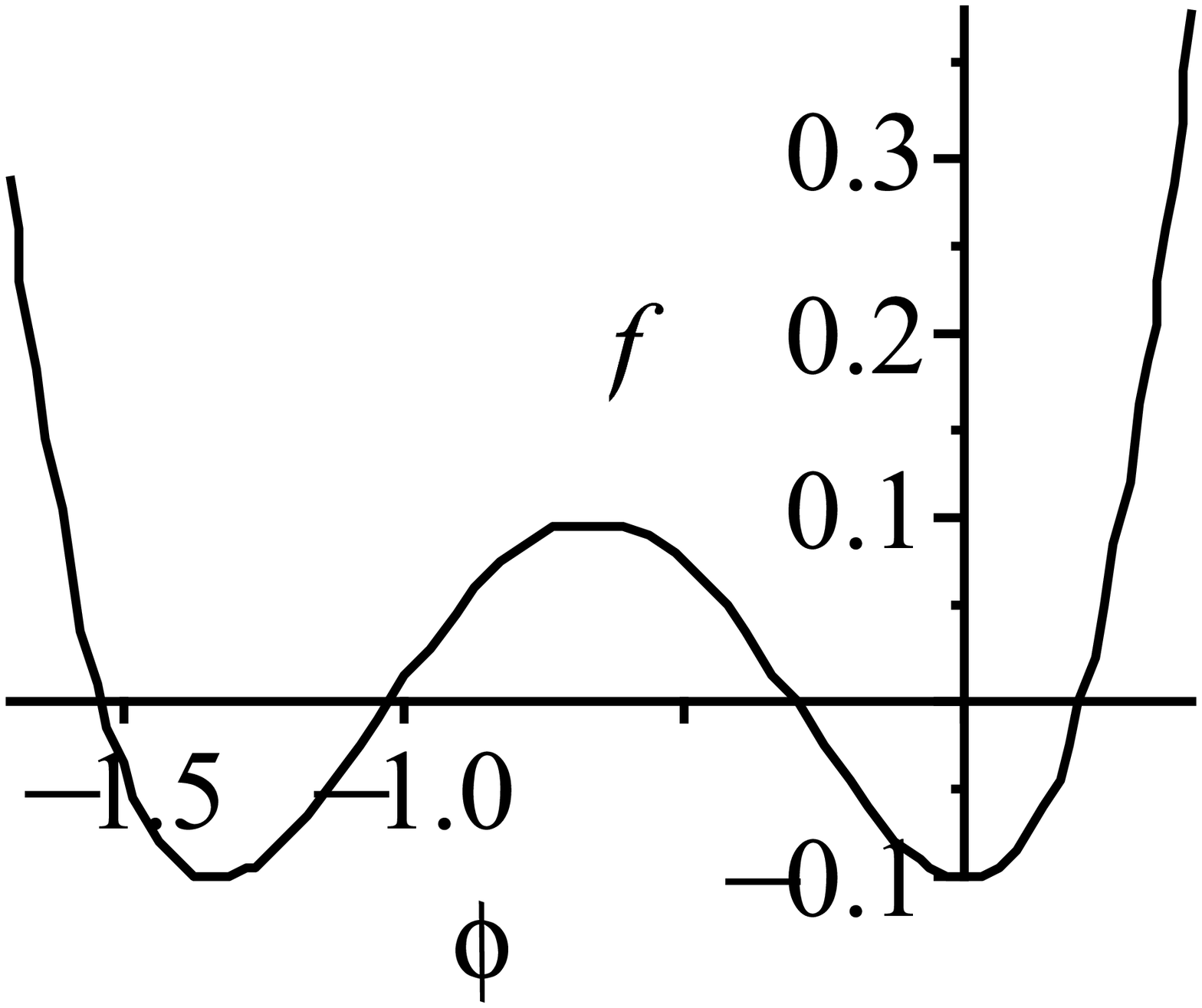}}\hspace{0.01\textwidth}
\subfloat[ ]{ \label{fig:6}
\includegraphics[height=1.in,width=1.2in]{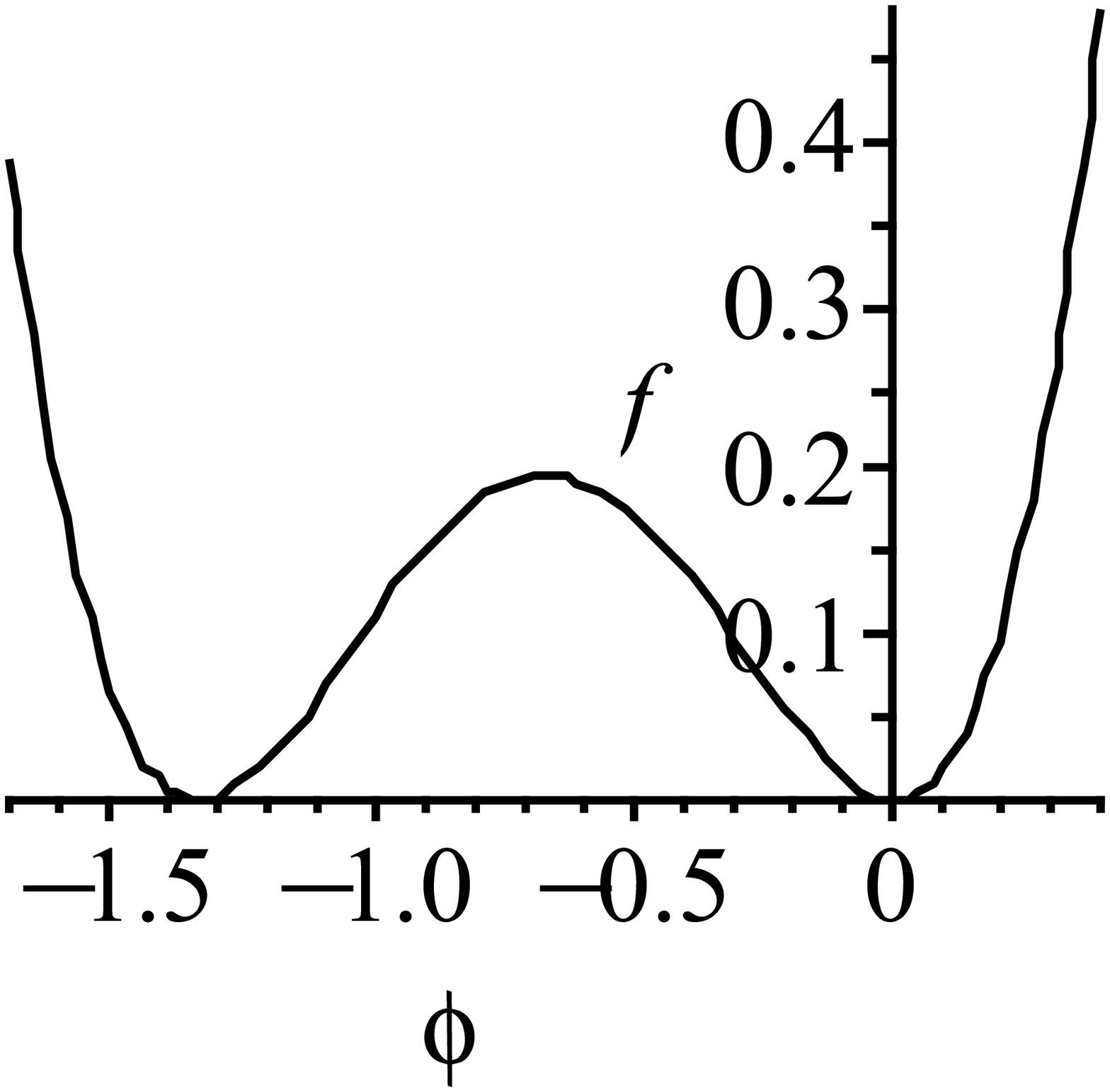}}\hspace{0.01\textwidth}
\subfloat[]{ \label{fig:7}
\includegraphics[height=1in,width=1.2in]{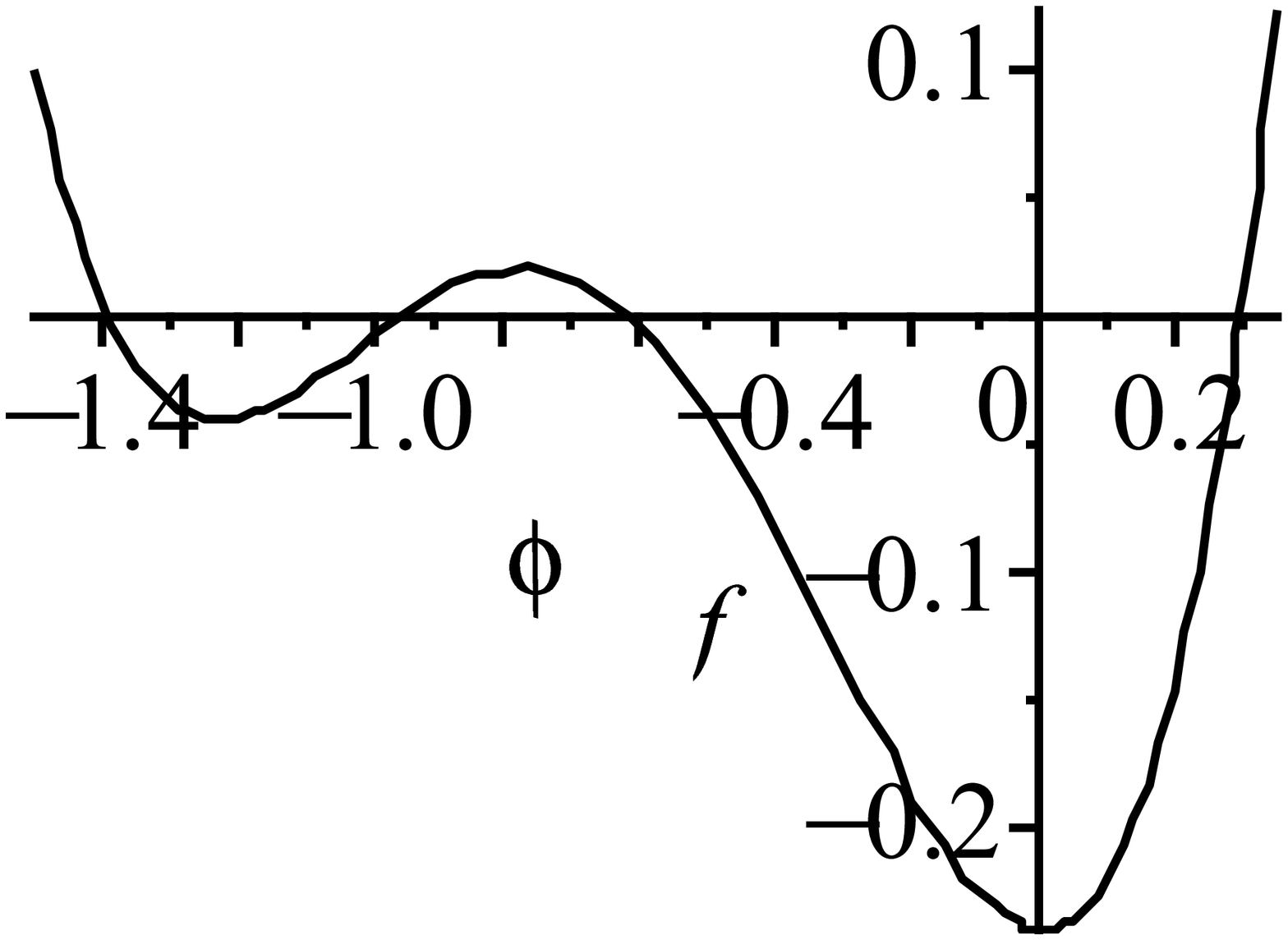}}\hspace{0.01\textwidth}
\subfloat[ ]{ \label{fig:8}
\includegraphics[height=1in,width=1.2in]{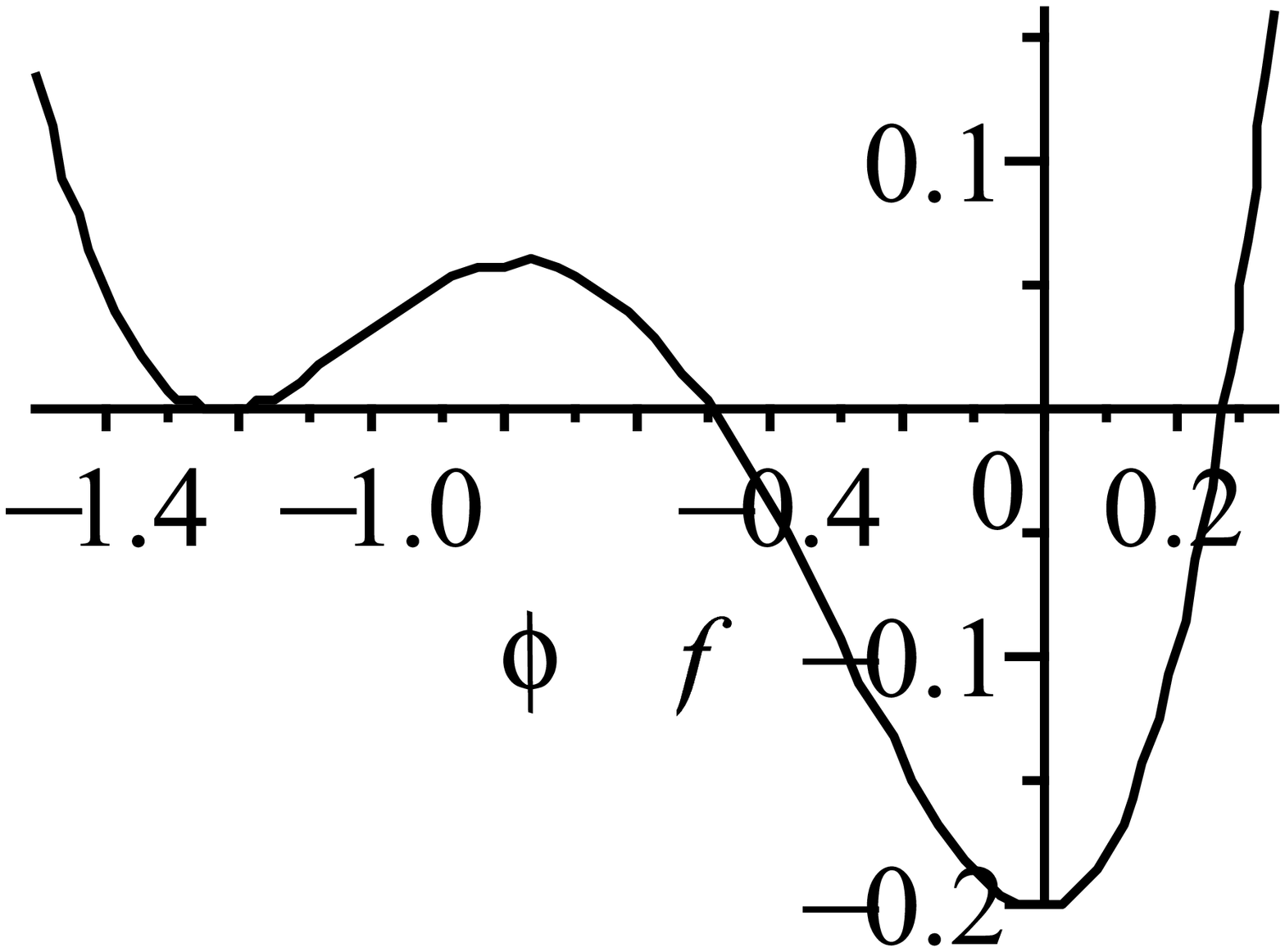}}

\caption{The curve of  $f(\varphi)$. (a) $a=-50$, $b=233$;
 (b) $a=-50$, $b=374.1346$;
 (c) $a=1.5$, $b=-0.05$;
 (d) $a=1.5$, $b=0$;
 (e) $a=\frac{\textstyle 16}{\textstyle 9}$, $b=-0.1$;
 (f) $a=\frac{\textstyle 16}{\textstyle 9}$, $b=0$;
 (g) $a=1.9$, $b=-0.24$;
 (h) $a=1.9$, $b=-0.2010$.}
\end{figure}

The case $a<0$ and $b=b_R$ (a corresponding curve of $f(\varphi)$ is
shown in Fig.1(b)) corresponds to the limit
$\varphi_3=\varphi_4=\varphi_R$ so that $m=1$, and then the solution
is an inverted loop-like solitary wave given by (\ref{eq2.9}) with
$\varphi_2\leq \varphi <\varphi_R$ and
\begin{equation}
  \label{eq3.9}
  \varphi_1=-\frac{\textstyle1}{\textstyle6}(2+3\sqrt{4-2a}+2\sqrt{4+6\sqrt{4-2a}}),
 \end{equation}
 \begin{equation}
  \label{eq3.10}
   \varphi_2=-\frac{\textstyle1}{\textstyle6}(2+3\sqrt{4-2a}-2\sqrt{4+6\sqrt{4-2a}}),
 \end{equation}
 see Fig.3(a) for
an example.

(2) $0<a<\frac{\textstyle 16}{\textstyle 9}$

In this case $\varphi_L<\varphi_R<0$ and $f(\varphi_L)<f(0)$. For
each value $0<a<\frac{\textstyle 16}{\textstyle 9}$ and $b_R<b<0$ (a
corresponding curve of $f(\varphi)$ is shown in Fig.1(c)), there are
periodic hump-like solutions to Eq.(\ref{eq3.3}) given by
(\ref{eq2.5}) so that $0<m<1$, and with wavelength given by
(\ref{eq2.8}); see Fig.2(b) for an example.

The case $0<a<\frac{\textstyle 16}{\textstyle 9}$ and $b=0$ (a
corresponding curve of $f(\varphi)$ is shown in Fig.1(d))
corresponds to the limit $\varphi_3=\varphi_4=0$ so that $m=1$, and
then the solution can be given by (\ref{eq2.9}) with $\varphi_1$ and
$\varphi_2$ given by the roots of $g(\varphi)=0$, namely
\begin{equation}
 \label{eq3.11} \varphi_1=-\frac{\textstyle
4}{\textstyle 3}-\frac{\textstyle 1}{\textstyle 3}\sqrt{16-9a},
\varphi_2=-\frac{\textstyle 4}{\textstyle 3}+\frac{\textstyle
1}{\textstyle 3}\sqrt{16-9a}.
 \end{equation}
In this case we obtain a weak solution, namely the periodic
upward-cusp wave
 \begin{equation}
  \label{eq3.12}
  \varphi=\varphi(\xi-2j\xi_m), (2j-1)\xi_m<\xi<(2j+1)\xi_m, \
  j=0,\pm1, \pm2, \cdots,
 \end{equation}
where
\begin{equation}
  \label{eq3.13}
  \varphi(\xi)=(\varphi_2-\varphi_1\tanh^2(\xi/4))\cosh^2(\xi/4),
 \end{equation}
and
\begin{equation}
  \label{eq3.14}
  \xi_m=4\tanh^{-1}\sqrt{\frac{\varphi_2}{\varphi_1}},
 \end{equation}
 see Fig.3(b) for an example.

 \begin{figure}[h]
\centering \subfloat[]{\label{fig:1}
\includegraphics[height=1.1in,width=2.4in]{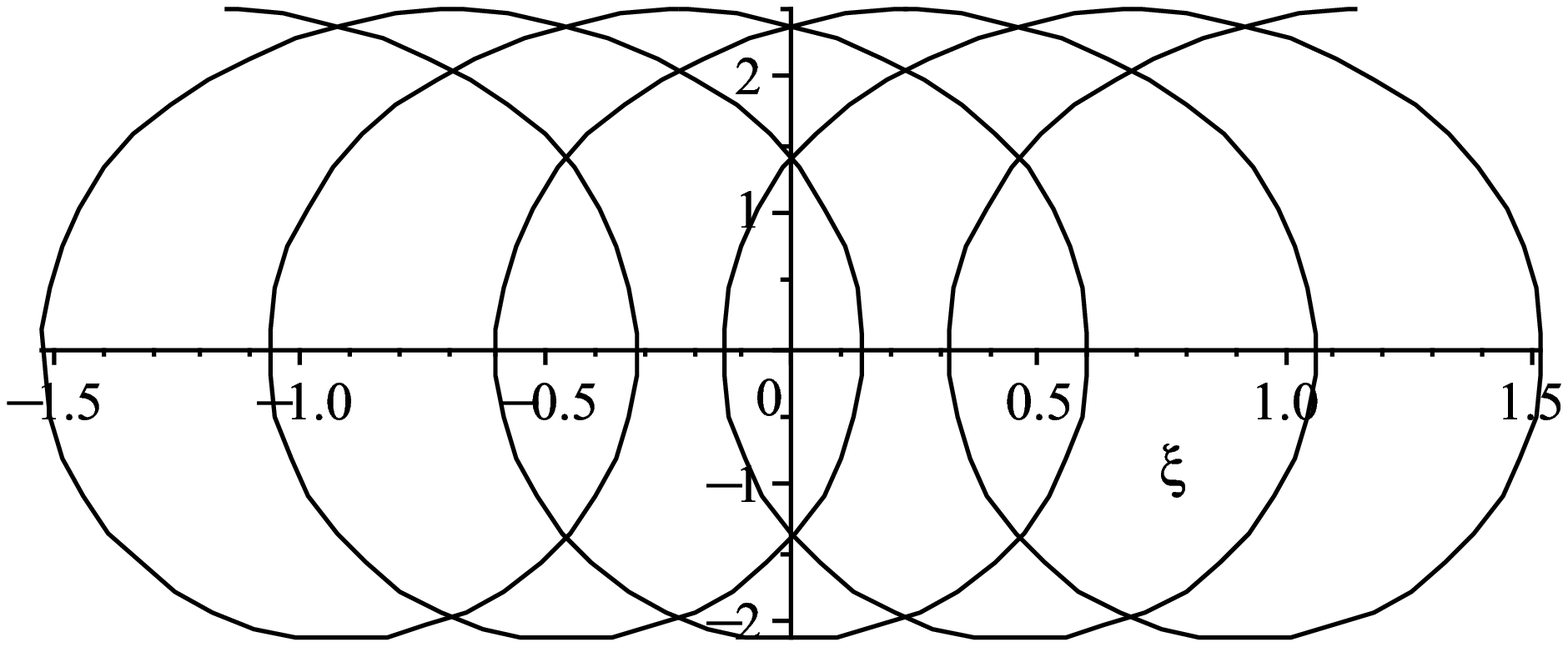}}\hspace{0.05\textwidth}
\subfloat[ ]{ \label{fig:2}
\includegraphics[height=1.2in,width=2.4in]{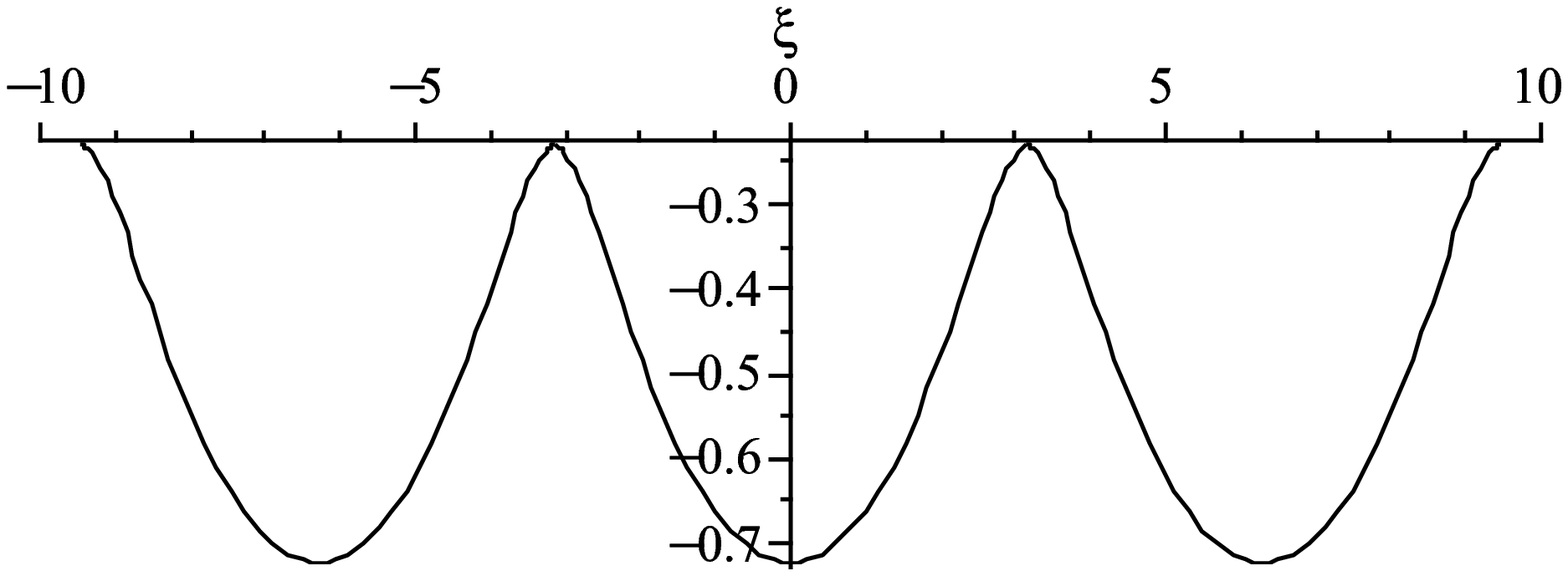}}\\
\subfloat[]{ \label{fig:3}
\includegraphics[height=1.2in,width=2.4in]{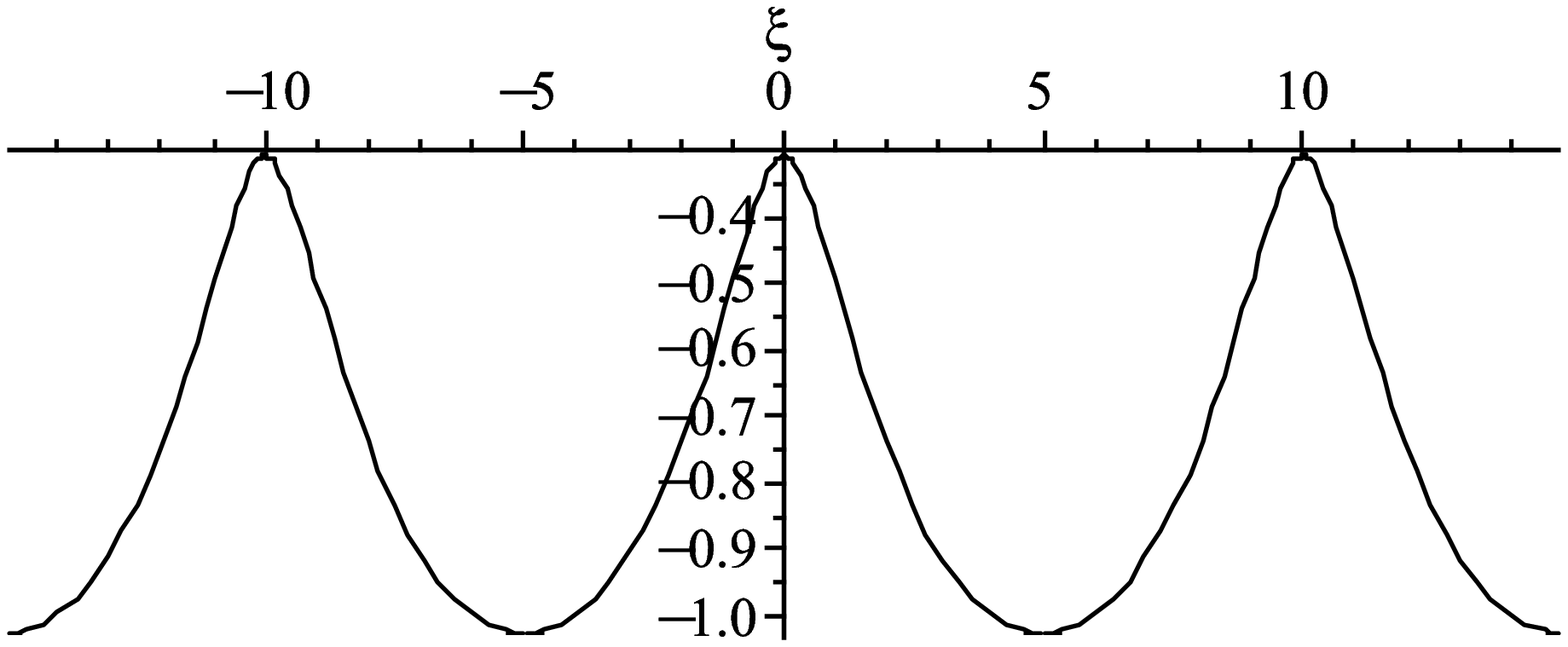}}\hspace{0.05\textwidth}
\subfloat[ ]{ \label{fig:4}
\includegraphics[height=1.2in,width=2.4in]{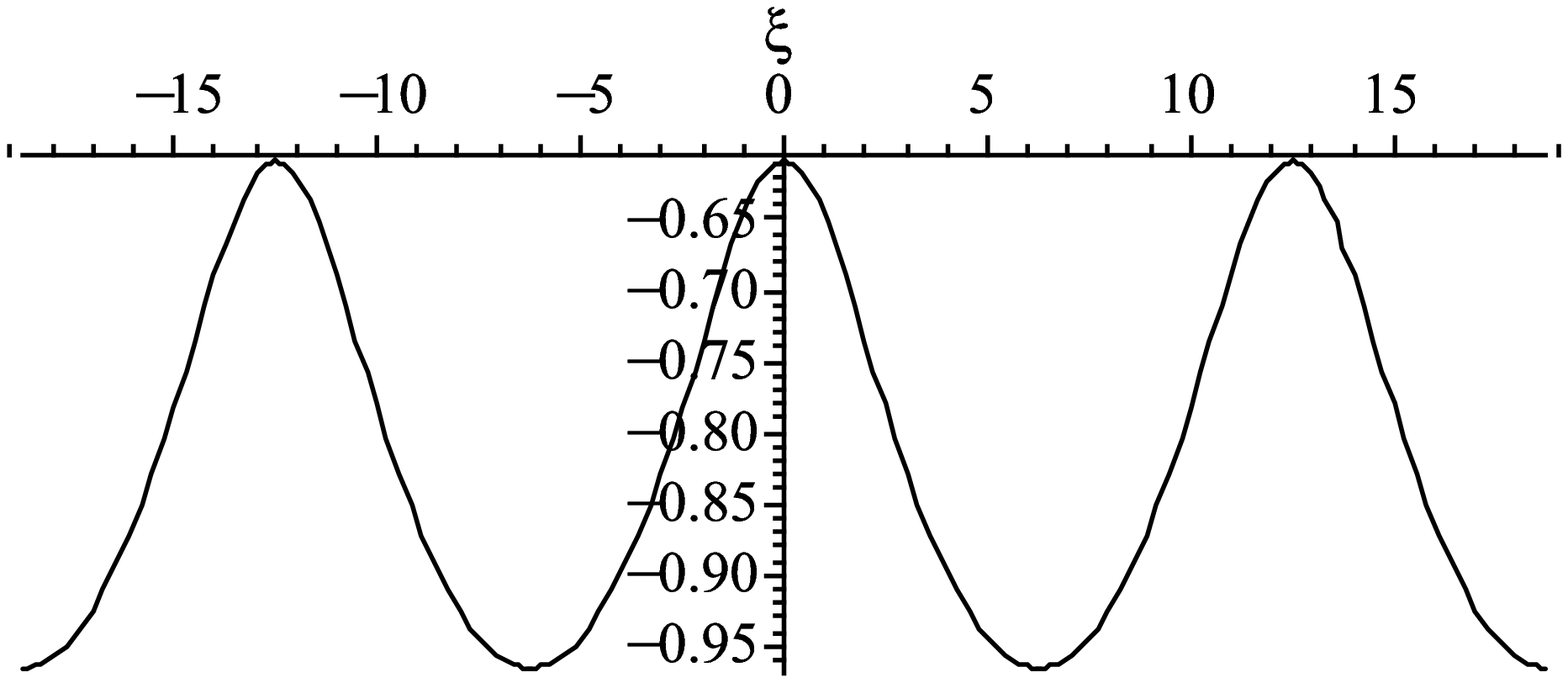}}

\caption{Periodic solutions to Eq.(\ref{eq3.3}) with $0<m<1$.
 (a) $a=-50$, $b=233$ so $m=0.7885$;
 (b) $a=1.5$, $b=-0.05$ so $m=0.6893$;
 (c) $a=\frac{\textstyle 16}{\textstyle 9}$, $b=-0.1$ so $m=0.8254$;
 (d) $a=1.9$, $b=-0.24$ so $m=0.6121$.}
\end{figure}

 \begin{figure}[h]
\centering \subfloat[]{\label{fig:1}
\includegraphics[height=1.2in,width=2.2in]{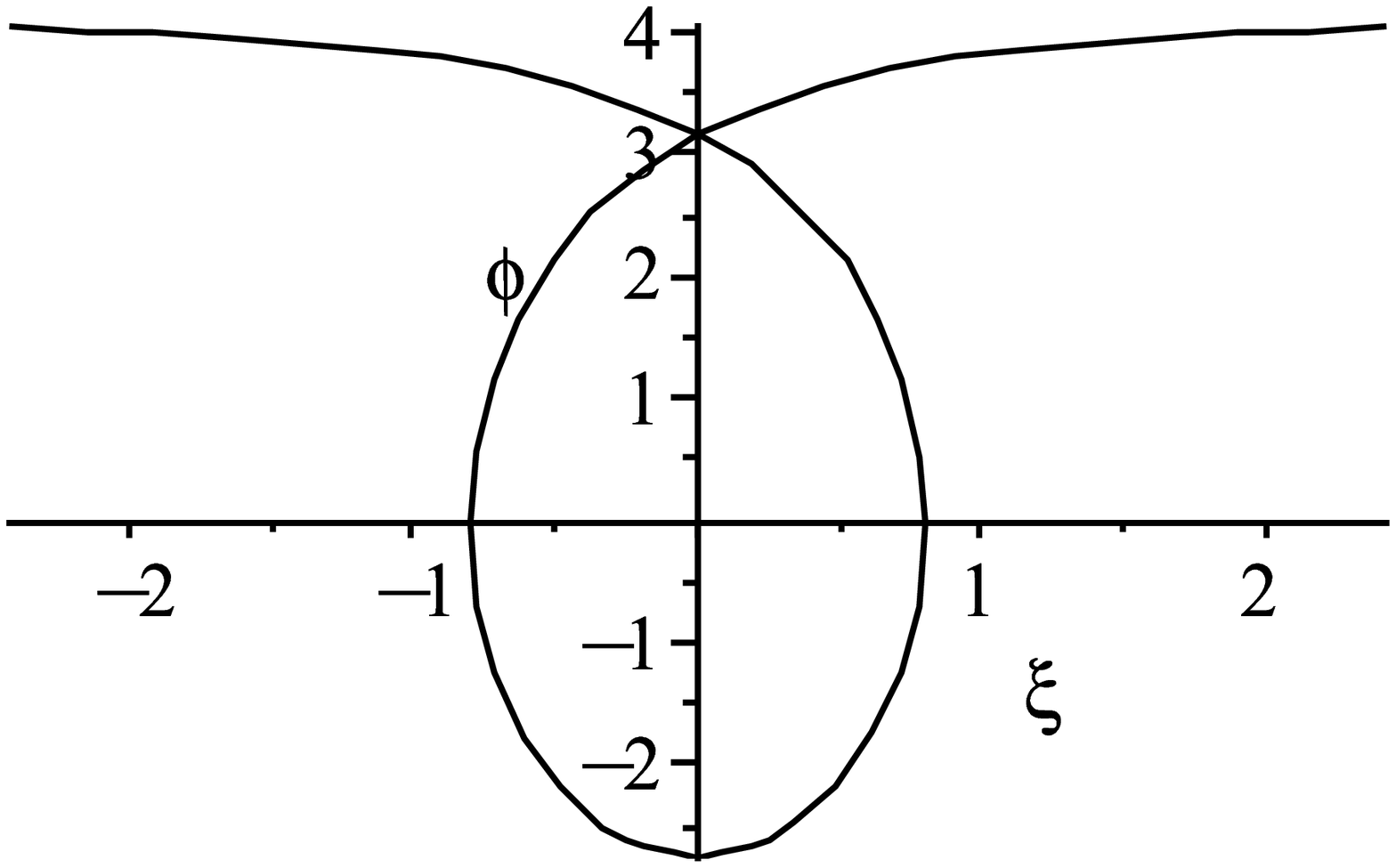}}\hspace{0.1\textwidth}
\subfloat[ ]{ \label{fig:2}
\includegraphics[height=1.2in,width=2.2in]{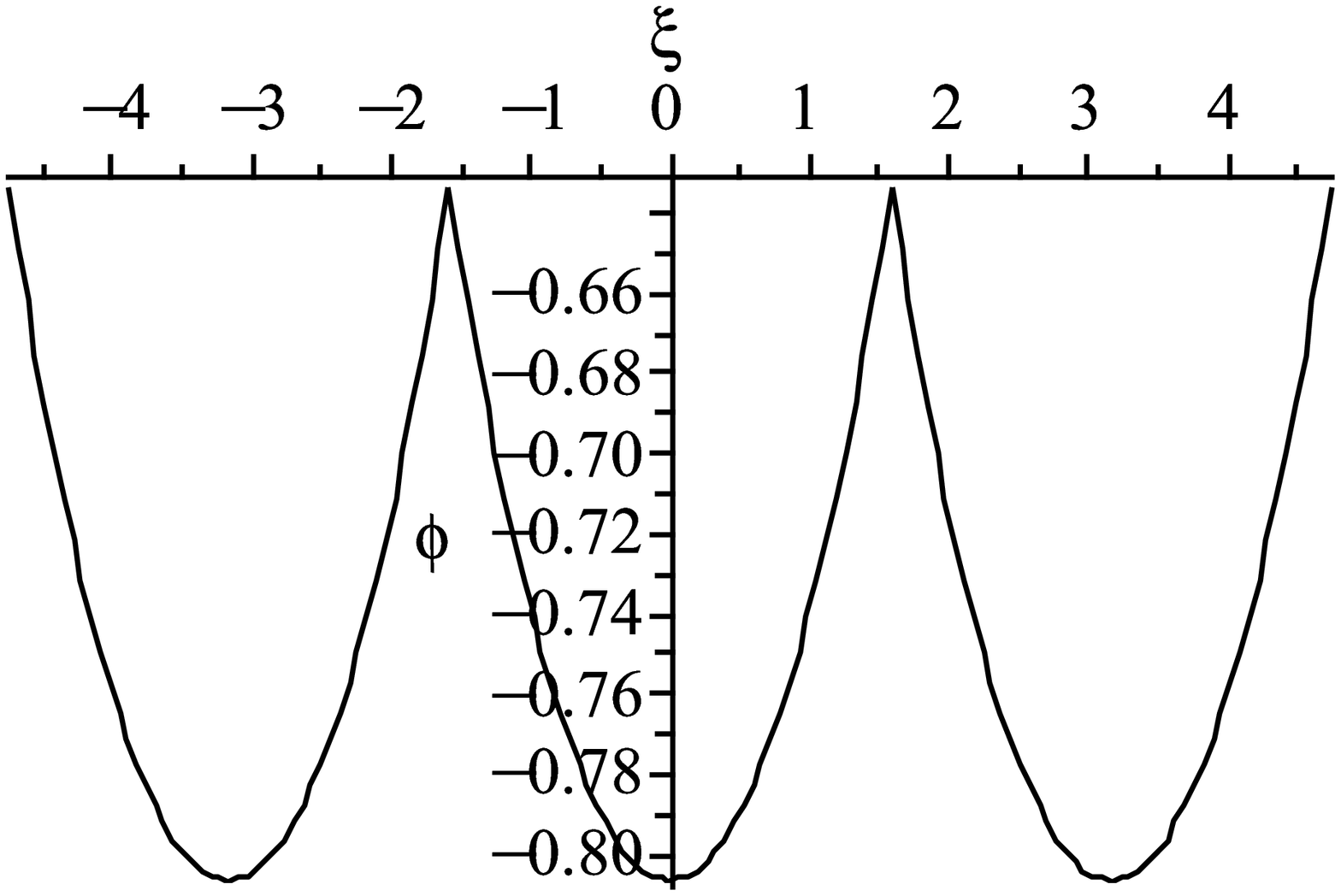}}\\
\subfloat[]{ \label{fig:3}
\includegraphics[height=1.2in,width=2.2in]{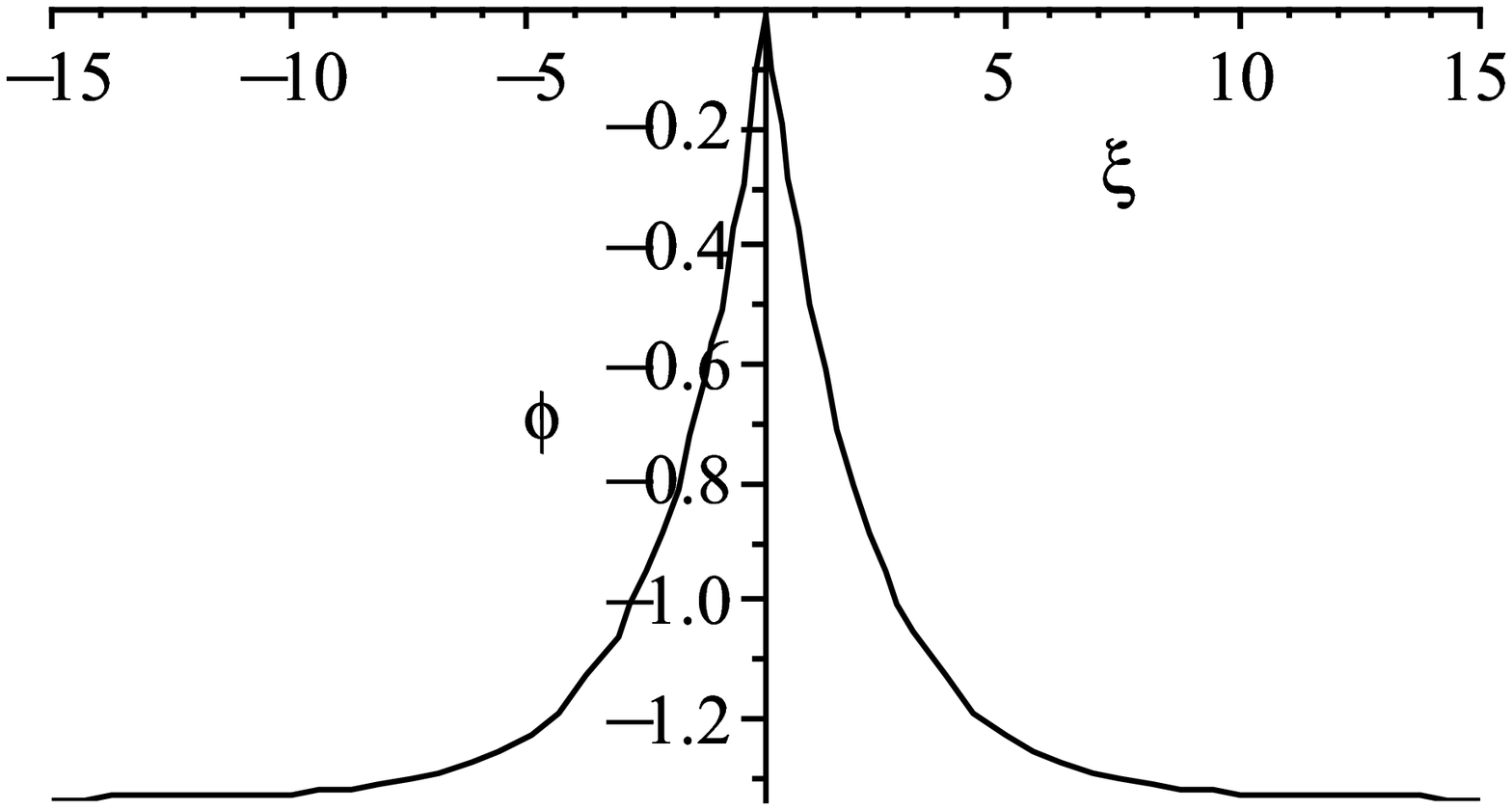}}\hspace{0.1\textwidth}
\subfloat[ ]{ \label{fig:4}
\includegraphics[height=1.2in,width=2.2in]{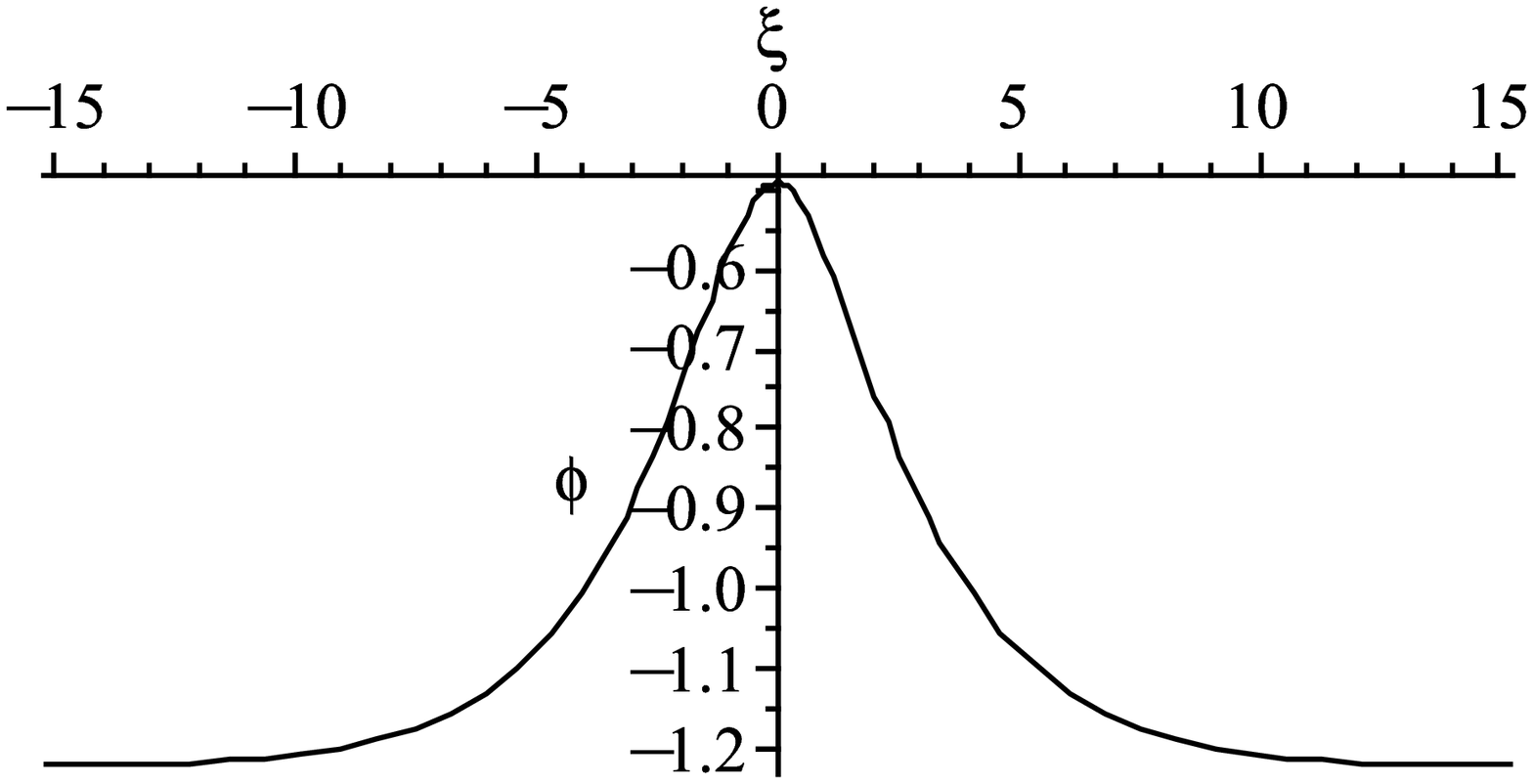}}
\caption{Solutions to Eq.(\ref{eq3.3}) with $m=1$.
 (a) $a=-50$, $b=374.1346$;
 (b) $a=1.5$, $b=0$;
 (c) $a=\frac{\textstyle 16}{\textstyle 9}$, $b=0$;
 (d) $a=1.9$, $b=-0.2010$.}
\end{figure}

 (3) $a=\frac{\textstyle 16}{\textstyle 9}$

In this case $\varphi_L<\varphi_R<0$ and $f(\varphi_L)=f(0)$.  For
$a=\frac{\textstyle 16}{\textstyle 9}$ and each value $b_R<b<0$ (a
corresponding curve of $f(\varphi)$ is shown in Fig.1(e)), there are
periodic hump-like solutions to Eq.(\ref{eq3.3}) given by
(\ref{eq2.10}) so that $0 < m < 1$, and with wavelength given by
(\ref{eq2.12}); see Fig.2(c) for an example.

The case $a=\frac{\textstyle 16}{\textstyle 9}$ and $b=0$ (a
corresponding curve of $f(\varphi)$ is shown in Fig.1(f))
corresponds to the limit
$\varphi_1=\varphi_2=\varphi_L=-\frac{\textstyle 4}{\textstyle 3}$
and $\varphi_3=\varphi_4=0$ so that $m=1$. In this case neither
(\ref{eq2.5}) nor (\ref{eq2.10}) is appropriate. Instead we consider
Eq.(\ref{eq3.3}) with $f(\varphi)=\frac{\textstyle 1}{\textstyle
4}\varphi^2(\varphi+\frac{\textstyle 4}{\textstyle 3})^2$ and note
that the bound solution has $-\frac{\textstyle 4}{\textstyle
3}<\varphi\leq 0$. On integrating Eq.(\ref{eq3.3}) and setting
$\varphi=0$ at $\xi=0$ we obtain a weak solution
\begin{equation}
  \label{eq3.15}
  \varphi=\frac{\textstyle 4}{\textstyle 3}\exp{(-\frac{\textstyle1}{\textstyle2}|\xi|)}-\frac{\textstyle 4}{\textstyle 3},
 \end{equation}
i.e. a single peakon solution with amplitude $\frac{\textstyle
4}{\textstyle 3}$, see Fig.3(c).

 (4) $\frac{\textstyle 16}{\textstyle 9}<a<2$

In this case $\varphi_L<\varphi_R<0$ and $f(\varphi_L)>f(0)$.  For
each value $\frac{\textstyle 16}{\textstyle 9}<a<2$ and $b_R<b<b_L$
(a corresponding curve of $f(\varphi)$ is shown in Fig.1(g)), there
are periodic hump-like solutions to Eq.(\ref{eq3.3}) given by
(\ref{eq2.10}) so that $0 < m < 1$, and with wavelength given by
(\ref{eq2.12}); see Fig.2(d) for an example.

The case $\frac{\textstyle 16}{\textstyle 9}<a<2$ and $b=b_L$ (a
corresponding curve of $f(\varphi)$ is shown in Fig.1(h))
corresponds to the limit $\varphi_1=\varphi_2=\varphi_L$ so that
$m=1$, and then the solution is a hump-like solitary wave given by
(\ref{eq2.13}) with $\varphi_L<\varphi\leq\varphi_3$ and
\begin{equation}
  \label{eq3.16}
  \varphi_3=\frac{\textstyle1}{\textstyle6}(-2+3\sqrt{4-2a}-2\sqrt{4-6\sqrt{4-2a}}),
 \end{equation}
 \begin{equation}
  \label{eq3.17}
   \varphi_4=\frac{\textstyle1}{\textstyle6}(-2+3\sqrt{4-2a}+2\sqrt{4-6\sqrt{4-2a}}),
 \end{equation}
see Fig.3(d) for an example.

On the above, we have obtained expressions of parametric form for
periodic and solitary wave solutions $\varphi(\xi)$ to
Eq.(\ref{eq3.3}). So in terms of $u=\varphi(\xi)+c$, we can get
expressions for the periodic and solitary wave solutions $u(\xi)$ to
Eq.(\ref{eq1.5}).
\section{Conclusion}

In this paper, we have found expressions for new travelling wave
solutions to the Fornberg-Whitham equation. These solutions depend,
in effect, on two parameters $a$ and $m$. For $m=1$, there are
inverted loop-like ($a<0$), single peaked
($a=\frac{\textstyle16}{\textstyle 9}$) and hump-like
($\frac{\textstyle 16}{\textstyle 9}<a<2$) solitary wave solutions.
For $m=1, 0<a<\frac{\textstyle 16}{\textstyle 9}$ or $0<m<1, a<2$
and $a \neq 0$, there are periodic hump-like wave solutions.

\end{document}